\documentstyle[12pt,aps]{revtex}

\widetext
\draft
\tighten

\begin{document}

\preprint{EFUAZ FT-99-73}

\title{{\rm {\bf The Modified Bargmann-Wigner Formalism:
Longitudinal Fields, Parity and All That}}\thanks{An invited paper for the
SSU-JINR collection on the occasion of the 90th anniversary
of the Saratov State University.}}

\author{{\bf Valeri V. Dvoeglazov}}

\address{{\rm
Escuela de F\'{\i}sica, Universidad Aut\'onoma de Zacatecas \\
Apartado Postal C-580, Zacatecas 98068, ZAC., M\'EXICO\\
E-mail:  valeri@ahobon.reduaz.mx\\
URL: http://ahobon.reduaz.mx/\~\,valeri/valeri.htm}}

\author{
and\\
{\bf Sergei V. Khudyakov}}

\address{\rm
Department of Theoretical and Nuclear Physics\\
Saratov State University\\
Astrakhanskaya ul., 83, Saratov 410071 RUSSIA}

\bigskip

\date{August 29, 1999}

\maketitle

\bigskip

\begin{abstract}
In the old papers of Ogievetski\u{\i} and Polubarinov, Hayashi, Kalb and
Ramond the {\it notoph} concept, the longitudinal field originated from
the antisymmetric tensor, has been proposed. In our work we analyze the
theory of antisymmetric tensor field of the second rank from a viewpoint
of the normalization problem. We obtain the 4-potentials and field
strengths, which coincide with those which have been previously obtained in
the works of Ahluwalia and Dvoeglazov. Slightly modifying the
Bargmann-Wigner field function we conclude that it is possible to describe
explicitly the degrees of freedom of the photon and of the {\it notoph}
by the same equation.  The physical consequences,
such as parity properties of field functions, are discussed, relations
to the previous works are discussed as well.  Moreover, we derive
equations for {\it symmetric } tensor of the second rank on the basis of
the same modification of the Bargmann-Wigner formalism, i.~e. the
equations which describe dynamical behavior of the fields of maximal spin
2.
\end{abstract}



The general outline for derivation of high-spin equations
has been given in~\cite{BW}.  A field of mass $m$ and spin
$j\geq {1\over 2}$ is represented by a totally-symmetric multispinor
of the rank $2j$. Particular cases
$j=1$ and $j={3\over 2} $ have been given in the textbooks, e.~g.,
ref.~\cite{Lurie}. In the present work we first prove that the
formalism can be generalized for the spin-1 and spin-3/2 cases.
The physical reason for these generalizations is that
the ordinary Bargmann-Wigner formalism does {\it not}
take into account the properties of field functions with respect to
discrete symmetry operations. The spin-2 case can also be of some
interest, because one can assume that the essential features of
the gravitational field (gravitons) are obtained from the transverse
components of the $(2,0)\oplus (0,2)$ representation of the Lorentz group.
Nevertheless, the questions of redundant components in the high-spin
relativistic equations are not yet understood in detail~\cite{Kirch}.

In a recent series of the papers~\cite{DVO0,DVO1,DVO2,DVA0,DVA1,DVA2},
we tried to construct
a consistent theory for the quantized {\it antisymmetric} tensor field
of the second rank, for the  4-vector field and for the {\it symmetric}
tensor field.
The previously published works~\cite{OP,HA,KR}, textbooks as well, cannot
be considered as the works that solved the main problems: Whether the
quantized AST field  and the quantized 4-potential field are `transverse'
fields or `longitudinal' ones (in the sense if the helicity $h=\pm 1$ or
$h= 0$)?  Is the  electromagnetic potential  a 4-vector in the quantized
theory?  How should the massless limit be taken?  Can the symmetric
tensor field of the second rank be consistently quantized? etc.  etc.
Many problems of the rigorous description of light
and  gravitation are still awaiting their solution and explicit
presentation.
Ideas of this article are based on three referee reports from`` Foundation
of Physics", even if they were critical.  First of all, we notice after the
referee that 1)``...In natural units ($c=\hbar= 1$) a Lagrangian density
has (because the action is considered to be dimensionless) dimension of
[energy]$^4$"; 2) One can always re-normalize the Lagrangian density and
``one can obtain the same equations of motion ...  by substituting
$L\rightarrow (1/M^N) L$, where $M$ is the arbitrary energy scale",
cf.~\cite{DVO1}; 3) the correct physical dimension  of the electromagnetic
field tensor $F^{\mu\nu}$ is [energy]$^2$; ``the transformation
$F^{\mu\nu}\rightarrow (1/2m) F^{\mu\nu}$ (see in ref.~[5,6a]) is required
a detailed study, because the mentioned transformation changes its
physical dimension:  it is not a simple normalization transformation."
Moreover, in the first articles on the {\it
notoph}~\cite{OP,HA,KR}\footnote{It is also known as a longitudinal field
of Kalb and Ramond, but the Ogievetski\u{\i} and Polubarinov consideration
seems to be more rigorous because it allows us to study the limit
$m\rightarrow 0$.} the authors used the normalization of the vector field
$F^\mu$\,\,\, \footnote{It is well known that the {\it notoph} field is
related to a field tensor of the third rank.} to [energy]$^2$ and,
therefore, the tensor potential $A^{\mu\nu}$, to [energy]$^1$.

Keeping in mind these observations, permit us
to repeat the derivation procedure for the Proca equations
from the equations of Bargmann and Wigner for a  totally
{\it symmetric} spinor of the second rank. We put
\begin{equation}
\Psi_{\{\alpha\beta\}} = (\gamma^\mu R)_{\alpha\beta} (c_a m A_\mu + c_f
F_\mu) +c_A m (\gamma^5 \sigma^{\mu\nu} R)_{\alpha\beta}
A_{\mu\nu} + c_F (\sigma^{\mu\nu} R)_{\alpha\beta}
F_{\mu\nu}\, ,\label{si} \end{equation}
where \begin{equation}
R=\pmatrix{i\Theta & 0\cr 0&-i\Theta\cr}\quad,\quad \Theta = -i\sigma_2 =
\pmatrix{0&-1\cr 1&0\cr}\, .
\end{equation}
Matrices $\gamma^{\mu} $ are chosen in the Weyl representation, i.~e.,
$\gamma^5$ is assumed to be diagonal.  The constants $c_i$ are some
numerical dimensionless coefficients. The reflection operator
$R$ has the following properties:
\begin{mathletters}
\begin{eqnarray}
&& R^T = -R\,,\quad R^\dagger =R = R^{-1}\,,\quad
R^{-1} \gamma^5 R = (\gamma^5)^T\,,\\
&& R^{-1}\gamma^\mu R = -(\gamma^\mu)^T\,,\quad
R^{-1} \sigma^{\mu\nu} R = - (\sigma^{\mu\nu})^T\,.
\end{eqnarray}
\end{mathletters}
that are necessary for the expansion (\ref{si})
to be possible in such a form, i.~e., $\gamma^{\mu} R$, $\sigma^{\mu\nu} R$
$\gamma^5\sigma^{\mu\nu} R$ are assumed to be  {\it symmetric}
matrices.

The substitution of the preceding expansion into the
Bargmann and Wigner set~\cite{Lurie}
\begin{mathletters}
\begin{eqnarray}
\left [ i\gamma^\mu
\partial_\mu -m \right ]_{\alpha\beta} \Psi_{\{\beta\gamma\}} (x) &=&
0\,,\label{bw1}\\
\left [ i\gamma^\mu \partial_\mu -m \right
]_{\gamma\beta} \Psi_{\{\alpha\beta\}} (x) &=& 0\,  \label{bw2}
\end{eqnarray}
\end{mathletters}
gives us the new equations of Proca:
\begin{mathletters}
\begin{eqnarray}
&& c_a m (\partial_\mu A_\nu - \partial_\nu A_\mu ) +
c_f (\partial_\mu F_\nu -\partial_\nu F_\mu ) =
ic_A m^2 \epsilon_{\alpha\beta\mu\nu} A^{\alpha\beta} +
2 m c_F F_{\mu\nu} \, \label{pr1} \\
&& c_a m^2 A_\mu + c_f m F_\mu =
i c_A m \epsilon_{\mu\nu\alpha\beta} \partial^\nu A^{\alpha\beta} +
2 c_F \partial^\nu F_{\mu\nu}\, . \label{pr2}
\end{eqnarray}
\end{mathletters}
In the case $c_a=1$, $c_F={1\over 2}$, $c_f=c_A=0$
they reduce to the ordinary Proca equations.\footnote{However, we noticed
that the division by $m$ in the first equation is {\it not} well-defined
operation in the case when somebody becomes interested
in the  $m\rightarrow 0$ limit later on.  Probably, in order
to avoid this dark point, one may wish to write the Dirac equation
in the form
$\left [ (i\gamma^\mu \partial_\mu)/ m-\openone\right ]\psi (x)= 0$,
the one that follows immediately in the
derivation of the  Dirac equation on the basis of the Ryder
relation~\cite{Ryder,DVA0} and the Wigner rules
for a {\it boost} of the field function from the
system with zero linear momentum.} In the generalized case
one obtains dynamical equations that connect the
photon, the {\it notoph} and their potentials.  Divergent parts
(in $m\rightarrow 0$) of field functions and of
dynamical variables should be removed by corresponding `gauge'
transformations  (either electrtomagnetic gauge transformations or
Kalb-Ramond gauge transformations).  It is well
known that the massless {\it notoph} field  turns out to be
a pure longitudinal field when one
keeps in mind $\partial_\mu A^{\mu\nu}= 0$.

Apart from these dynamical equations, we can obtain a set of
constraints by means of subtraction of the equations of
Bargmann and Wigner (instead of their addition as in (\ref{pr1},\ref{pr2})).
They are read
\begin{equation} mc_a
\partial^\mu A_\mu + c_f \partial^\mu F_\mu =0,\, \quad\mbox{and} \quad
mc_A \partial^\alpha A_{\alpha\mu} + {i\over 2} c_F
\epsilon_{\alpha\beta\nu\mu} \partial^\alpha F^{\beta\nu} = 0\, .
\end{equation}
that suggest
$\widetilde F^{\mu\nu} \sim im A^{\mu\nu}$ $ F^\mu \sim m
A^\mu$, like in~\cite{OP}.

We returned to these old works
due to recent interpretational controversies after
experimental observations of the objects ${\bf E}\times
{\bf E}^\ast$, ${\bf B}^{(1)}\times {\bf B}^{(2)}$ and ${\bf A}\times {\bf
A}^\ast$ in nonlinear optics. In this respect one can consider that $\sim
{\bf A} \times {\bf A}^\ast$ can be a part of the tensor {\it potential}
and $\sim{\bf B}\times {\bf B}^\ast$, a part of the vector field (cf.  the
formulas (19a-c) in ref.~[6a]).\footnote{Recently we proved
(physics/9907048) that
such  cross products can have {\it various} transformations properties
with respect to the Lorentz transformations on the corresponding
transverse fields. The physical origin is the possibility of construction
of {\it various} field operators, differing by properties with
respect to the $C$ and $CP$ discrete operations.}
Following~\cite[Eqs.(9,10)]{OP}
we proceed in the construction of the ``potentials" for the
{\it notoph}: $A_{\mu\nu} ({\bf p})\sim\left [\epsilon_\mu^{(1)} ({\bf p})
\epsilon_\nu^{(2)} ({\bf p})- \epsilon_\nu^{(1)} ({\bf
p})\epsilon_\mu^{(2)} ({\bf p})\right ]$ upon using explicit forms of the
polarization vectors in the linear momentum space (e.~g.,
refs.~\cite{Wein} and~[6a, formulas (15a, b)]).  One obtains
\begin{eqnarray} A^{\mu\nu} ({\bf p}) = {iN^2 \over
m} \pmatrix{0&-p_2& p_1& 0\cr p_2 &0& m+{p_r p_l\over p_0+m} & {p_2
p_3\over p_0 +m}\cr -p_1 & -m - {p_r p_l \over p_0+m}& 0& -{p_1 p_3\over
p_0 +m}\cr 0& -{p_2 p_3 \over p_0 +m} & {p_1 p_3 \over p_0+m}&0\cr}\, ,
\label{lc} \end{eqnarray}
which coincides with the `longitudinal' components
of the  antisymmetric tensor which have been obtained in
refs.~\cite[Eqs. (2.14,2.17)]{DVA1} and in~[6a, Eqs. (17b,18b)]
within normalization and different forms of the
spinorial basis.  The longitudinal states can be eliminated in
the massless case when one adapted suitable
normalization and if a $j=1$ particle moves
along the third-axis $OZ$ direction. It is also useful to compare Eq.
(\ref{lc}) with the formula (B2) in ref.~\cite{DVA2}, the expressions
for the strengths in the light-front form of the QFT, in
order to realize the correct procedure for taking the massless limit.

As a discussion we want to mention that the Tam-Happer experiment~\cite{TH}
did not find a satisfactory explanation in the
quantumelectrodynamic frameworks
(at least, its explanation is complicated
by tedious technical calculus). On the other hand, in
ref.~\cite{Prad} an interesting
model has been proposed.  It is based on gauging the Dirac field
on using a set of parameters which are dependent on
space-time coordinates $\alpha_{\mu\nu} (x) $ in $\psi (x)\rightarrow
\psi^\prime (x^\prime)=\Omega\psi (x),\,\,\quad\Omega=\exp\left [
{i\over 2}\sigma^{\mu\nu}\alpha_{\mu\nu} (x)\right ]\, $.
Thus, the second ``photon" has been taken into consideration.
The 24-component compensation field  $B_{\mu,\nu\lambda} $
reduces to the 4-vector field as follows
(the notation of~\cite{Prad} is used here):  $B_{\mu,\nu\lambda}=
{1\over 4}\epsilon_{\mu\nu\lambda\sigma} a_\sigma (x)\, .$
As you can readily see after the
comparison of the formulas of~\cite{Prad} with those of
refs.~\cite{OP,HA,KR}, the second photon of Pradhan and Naik is nothing
more than the Ogievetski\u{\i}-Polubarinov {\it notoph} within the
normalization.  Parity properties (massless behavior as well) are the
matter of dependence, not only on the explicit forms  of the field
functions in the momentum space $ (1/2,1/2)$ representation, {\it but}
also on the properties of the corresponding creation/annihilation
operators.  The helicity properties  in the massless limit depend on the
normalization.

We can generalize the formalism in a slightly different way.
In the works~\cite{ras}  generalizations of the
Dirac formalism have been proposed. One of them has the form
\begin{equation}
\left [
i\gamma^\mu \partial_\mu - m_1 - \gamma^5 m_2 \right ] \Psi (x) =0\,
\label{rs}
\end{equation}
with two mass parameters and they allow the description of  tachyonic
states and, probably, the explanation of the mass hierarchy~\cite{Barut}.
If one wants to build the formalism for high spins on the basis of the
equation (\ref{rs}), one comes to the following system of Proca equations
for spin 1:\footnote{ For the spin- 3/2 case the equations are very
similar. It is only necessary to consider that the spin-3/2 function takes
an additional index which is related to its spinorial part.  However, new
restrictions, thanks to the symmetrization procedure of the third-rank
multispinor, appear (see ref.~\cite{Lurie}).}
\begin{mathletters}
\begin{eqnarray} && 2c_1 \partial_\mu \widetilde F^{\mu\alpha} -2ic_2
\partial_\mu A^{\mu\alpha}  + m_2 \Psi^\alpha =0\,,\\ && 2c_1 \partial_\mu
F^{\mu\alpha} +2ic_2 \partial_\mu \widetilde A^{\mu\alpha} + m_1
\Psi^\alpha =0\,,\\ && 2c_1 (m_1 F^{\mu\nu} + im_2 \widetilde F^{\mu\nu})
+2c_2 (m_2 A^{\mu\nu} +i m_1 \widetilde A^{\mu\nu}) - (\partial^\mu
\Psi^\nu - \partial^\nu \Psi^\mu ) = 0\, ,\\ && \partial_\mu \Psi^\mu =0\,
.  \end{eqnarray} \end{mathletters}
The field  function was assumed to be
represented in this case by \begin{equation} \Psi_{\{\alpha\beta\}} =
(\gamma^\mu R)_{\alpha\beta} \Psi_\mu + c_1 (\sigma^{\mu\nu}
R)_{\alpha\beta} F_{\mu\nu} +c_2 (\gamma^5 \sigma^{\mu\nu}
R)_{\alpha\beta} A_{\mu\nu}\,.  \end{equation} The use of two mass
parameters can also be useful for the consideration of the Pauli-Lubanski
vector  and the study of its massless limit (see~[6a, Eq.  (27,28)]).
The latter is interesting to a reader because the choices $m_1 =\pm m_2$
provide us with alternative formulation for massless particles $p^\mu
p_\mu =0$, even though a theory contains a parameter with dimension of
mass.

Furthermore, upon repeating the generalized formalism for a fourth-rank
{\it symmetric} multispinor
\begin{eqnarray}
&&\Psi_{\{\alpha\beta\}\{\gamma\delta\}} =
\alpha_1 \beta_1 (\gamma_\mu R)_{\alpha\beta} (\gamma^\kappa
R)_{\gamma\delta} G_\kappa^{\quad\mu} +\alpha_1 \beta_2
(\gamma_\mu R)_{\alpha\beta} (\sigma^{\kappa\tau} R)_{\gamma\delta}
F_{\kappa\tau}^{\quad\mu} + \nonumber\\
&+&\alpha_1 \beta_3 (\gamma_\mu R)_{\alpha\beta}
(\gamma^5 \sigma^{\kappa\tau} R)_{\gamma\delta} \widetilde
F_{\kappa\tau}^{\quad\mu} +
\alpha_2 \beta_4 (\sigma_{\mu\nu}
R)_{\alpha\beta} (\gamma^\kappa R)_{\gamma\delta} T_\kappa^{\quad\mu\nu}
+\nonumber\\
&+&\alpha_2 \beta_5 (\sigma_{\mu\nu} R)_{\alpha\beta} (\sigma^{\kappa\tau}
R)_{\gamma\delta} R_{\kappa\tau}^{\quad \mu\nu}
+ \alpha_2
\beta_6 (\sigma_{\mu\nu} R)_{\alpha\beta} (\gamma^5 \sigma^{\kappa\tau}
R)_{\gamma\delta} \widetilde R_{\kappa\tau}^{\quad\mu\nu} +\nonumber\\
&+&\alpha_3 \beta_7 (\gamma^5 \sigma_{\mu\nu} R)_{\alpha\beta}
(\gamma^\kappa R)_{\gamma\delta} \widetilde
T_\kappa^{\quad\mu\nu}+
\alpha_3 \beta_8 (\gamma^5
\sigma_{\mu\nu} R)_{\alpha\beta} (\sigma^{\kappa\tau} R)_{\gamma\delta}
\widetilde D_{\kappa\tau}^{\quad\mu\nu} +\nonumber\\
&+&\alpha_3 \beta_9
(\gamma^5 \sigma_{\mu\nu} R)_{\alpha\beta} (\gamma^5 \sigma^{\kappa\tau}
R)_{\gamma\delta} D_{\kappa\tau}^{\quad \mu\nu}\, .
\label{ffn1}
\end{eqnarray}
one  obtains the following
dynamical equations
\begin{mathletters} \begin{eqnarray} &&
{2\alpha_2 \beta_4 \over m} \partial_\nu T_\kappa^{\quad\mu\nu}
+{i\alpha_3 \beta_7 \over m} \epsilon^{\mu\nu\alpha\beta} \partial_\nu
\widetilde T_{\kappa,\alpha\beta} = \alpha_1 \beta_1
G_\kappa^{\quad\mu}\,, \label{b}\\ &&{2\alpha_2 \beta_5 \over m}
\partial_\nu R_{\kappa\tau}^{\quad\mu\nu} +{i\alpha_2 \beta_6 \over m}
\epsilon_{\alpha\beta\kappa\tau} \partial_\nu \widetilde R^{\alpha\beta,
\mu\nu} +{i\alpha_3 \beta_8 \over m}
\epsilon^{\mu\nu\alpha\beta}\partial_\nu \widetilde
D_{\kappa\tau,\alpha\beta} - \nonumber\\
&-&{\alpha_3 \beta_9 \over 2}
\epsilon^{\mu\nu\alpha\beta} \epsilon_{\lambda\delta\kappa\tau}
D^{\lambda\delta}_{\quad \alpha\beta} = \alpha_1 \beta_2
F_{\kappa\tau}^{\quad\mu} + {i\alpha_1 \beta_3 \over 2}
\epsilon_{\alpha\beta\kappa\tau} \widetilde F^{\alpha\beta,\mu}\,, \\
&& 2\alpha_2 \beta_4 T_\kappa^{\quad\mu\nu} +i\alpha_3 \beta_7
\epsilon^{\alpha\beta\mu\nu} \widetilde T_{\kappa,\alpha\beta}
=  {\alpha_1 \beta_1 \over m} (\partial^\mu G_\kappa^{\quad \nu}
- \partial^\nu G_\kappa^{\quad\mu})\,, \\
&& 2\alpha_2 \beta_5 R_{\kappa\tau}^{\quad\mu\nu} +i\alpha_3 \beta_8
\epsilon^{\alpha\beta\mu\nu} \widetilde D_{\kappa\tau,\alpha\beta}
+i\alpha_2 \beta_6 \epsilon_{\alpha\beta\kappa\tau} \widetilde
R^{\alpha\beta,\mu\nu}
- {\alpha_3 \beta_9\over 2} \epsilon^{\alpha\beta\mu\nu}
\epsilon_{\lambda\delta\kappa\tau} D^{\lambda\delta}_{\quad \alpha\beta}
= \nonumber\\
&=& {\alpha_1 \beta_2 \over m} (\partial^\mu F_{\kappa\tau}^{\quad \nu}
-\partial^\nu F_{\kappa\tau}^{\quad\mu} ) + {i\alpha_1 \beta_3 \over 2m}
\epsilon_{\alpha\beta\kappa\tau} (\partial^\mu \widetilde
F^{\alpha\beta,\nu} - \partial^\nu \widetilde F^{\alpha\beta,\mu} )\, .
\label{f}
\end{eqnarray}
\end{mathletters}
The essential constraints are given in Appendix B.
They are the results of contractions of the field function
(4-rank symmetric multispinor) with three antisymmetric matrices
$R^{-1}, R^{-1} \gamma^5, R^{-1}\gamma^5 \gamma^\mu$, as
has been done previously in the standard formalism (see~\cite{Lurie} for
the case of $j=3/2$).

Moreover, one obtains the second-order equation for the
second-rank symmetric tensor from a set of equations (\ref{b}-\ref{f})
($\alpha_1\neq 0$, $\beta_1\neq 0$):
\begin{equation} {1\over m^2} \left [\partial_\nu \partial^\mu
G_\kappa^{\quad \nu} - \partial_\nu \partial^\nu G_\kappa^{\quad\mu}
\right ] =  G_\kappa^{\quad \mu}\, .  \end{equation}
After contraction
in  indices $\kappa$ and $\mu$, this equation
reduces  to a set
\begin{eqnarray} \cases{{1\over m^2}
\partial_\mu G_{\quad\nu}^{\mu} = F_\nu\,  &\cr  \partial_\nu
F^\nu = 0&}\, , \end{eqnarray}
i.~e., to the equations that connect
an analog of the metric tensor and
an analog of the 4-potential.\footnote{Please note that the physical
dimension of the $F^\mu$ field in this case should be [energy]$^3$
if one considers $G^{\mu\nu}$ as the energy-momentum tensor.}
The gravitation theory based on these dynamical equations is certainly
related to the Logunov's relativistic theory of gravitation (RTG),
cf.~\cite[\S 36]{Logun}. Furthermore, in the linearized version of
gravitation theory (which may follow from the presented formalism)
-- in fact, in a weak limit -- only certain chiral components of the
4-rank symmetric multispinor contribute, as discussed in~\cite{Marques}.

In conclusion, we can say that in a series of work it was firmly
established: 1) the
physical significance of the normalization has been proved; 2) the {\it
notoph} concept, i.~e. the formalisms for the
Ogievetski\u{\i}-Polubarinov-Kalb-Ramond longitudinal field,  the torsion
field, the gauge `string' field etc.  are all connected to each other  and
are related to the helicity-0 component of the antisymmetric tensor field
of the second rank; 3) we proved that one can describe the `transverse' and
`longitudinal' degrees of freedom by the same equation with unknown
numerical coefficients~[6b]; 4) the dependence of coefficient functions in
the field operators for higher spins on rotations of the frame of
reference suggest interpretation of higher-spin particles as composite
particles; 5) the same procedure as for fields of spin-0 and spin-1 has
been applied to the case of spin 2 (and, presumably, it should be applied
to the cases of higher spins) for the reason that the standard formalism
has problems with interpretation of the existence of the trivial solutions
only~[6c]; 6) relative intrinsic parities of various tensors\footnote{This
is obviously measurable quantity.} considered in the present paper are
different and this fact forbids the identification of the physical
contents of theories containing/not containing dual tensors.

\bigskip

{\it Acknowledgments.} We are very thankful
to  colleagues over the world for the Internet communications.
Particularly, We want to mention discussions with Drs.
D. V. Ahluwalia, A. E.  Chubykalo, M.  Israelit, M. Kirchbach and A. S.
Shumovsky~\cite{Shumovsky}.   We acknowledge the technical help of
Mr. Jes\'us C\'azares Montes as well.
The work has been partially supported by the National System of
Investigators of M\'exico.

\bigskip

{\bf Appendix A.} Field functions of the $(1/2,1/2)$ representation in the
linear momentum space are derived to be (see ref.~[6a]):
\begin{mathletters} \begin{eqnarray} u^\mu ({\bf p}, +1)= -{N\over
m\sqrt{2}}\pmatrix{p_r\cr m+ {p_1 p_r \over p_0+m}\cr im +{p_2 p_r \over
p_0+m}\cr {p_3 p_r \over p_0+m}\cr}&\quad&,\quad u^\mu ({\bf p}, -1)=
{N\over m \sqrt{2}}\pmatrix{p_l\cr m+ {p_1 p_l \over p_0+m}\cr -im +{p_2
p_l \over p_0+m}\cr {p_3 p_l \over p_0+m}\cr}\quad,\quad\\ u^\mu ({\bf p},
0) = {N\over m} \pmatrix{p_3\cr {p_1 p_3 \over p_0+m}\cr {p_2 p_3 \over
p_0+m}\cr m + {p_3^2 \over p_0+m}\cr}&\quad&, \quad u^\mu ({\bf p}, 0_t) =
{N \over m} \pmatrix{p_0\cr p_1 \cr p_2\cr p_3\cr}\quad.  \end{eqnarray}
\end{mathletters}
They are not divergent in the massless limit
if we take $N= m$. They describe the longitudinal ``photons"
(in the sense that $h=0$) in this limit.  If $N=1$ one has
``transverse photons" but we also have  the divergent conduct
of the gauge part of  the field functions~\cite[\S
3.2.3]{Itzyk}.\footnote{This fact induced someone to consider additional
{\it scalar} field (i.~e. $j=0$ field, sic!) in order to make the spin-1
formalism to be consistent.  It is the Stueckelberg formulation.} The
potentials for negative-energy solutions  are obtained by application of
the complex-conjugation operator (in fact, the charge conjugation operator
in this representation) or the $CP$ operator.

The physical strengths are
\begin{mathletters}
\begin{eqnarray}
{\bf B}^{(+)} ({\bf p}, +1) &=& -{iN\over 2\sqrt{2} m} \pmatrix{-ip_3 \cr
p_3 \cr ip_r\cr}\,,\quad
{\bf E}^{(+)} ({\bf p}, +1) =  -{iN\over 2\sqrt{2} m} \pmatrix{p_0- {p_1
p_r \over p_0+m}\cr ip_0 -{p_2 p_r \over p_0+m}\cr -{p_3 p_r \over
p_0+m}\cr} \,,\\
{\bf B}^{(+)} ({\bf p}, 0) &=& {iN \over 2m}
\pmatrix{p_2 \cr -p_1 \cr 0\cr}\,,\quad
{\bf E}^{(+)} ({\bf p}, 0) =  {iN \over 2m} \pmatrix{- {p_1 p_3
\over p_0+m}\cr -{p_2 p_3 \over p_0+m}\cr p_0-{p_3^2 \over
p_0+m}\cr} \,,\\
{\bf B}^{(+)} ({\bf p},
-1) &=& {iN \over 2\sqrt{2}m} \pmatrix{ip_3 \cr p_3 \cr
-ip_l\cr} \,,\quad
{\bf E}^{(+)} ({\bf p}, -1) =  {iN\over 2\sqrt{2} m} \pmatrix{p_0- {p_1
p_l \over p_0+m}\cr -ip_0 -{p_2 p_l \over p_0+m}\cr -{p_3 p_l \over
p_0+m}\cr} \,.
\end{eqnarray} \end{mathletters}
They are obtained by the application of  formulas:
${\bf B}^{(\pm)} ({\bf p}, h)=\pm{ i\over 2m}{\bf p}\times {\bf u}^{(\pm)}
({\bf p}, h) $ and ${\bf E}^{(\pm)} ({\bf p}, h)=\pm
{ i\over 2m} p_0 {\bf u}^{(\pm)} ({\bf p}, h)\mp { i\over 2m}{\bf p}
u^{0\, (\pm)} ({\bf p}, h) $.  It is useful to compare
these  formulas with those that have been presented in~\cite[p.408]{DVA1},
with taking into account that the author~\cite{DVA1}
used a  different spin basis
therein.  Cross products also have been obtained in ref.~[6a] and, as we
can see, they are related to the gauge part of ($\sim p^\mu$) of
the 4-potential  in the momentum space.

\bigskip

{\bf Appendix B.}
The constraints for the spin-2 case  have been
obtained in~[6c]. They are the results of contraction of
the 4-rank symmetric multispinor with corresponding
antisymmetric matrices of the set of $\Gamma R$-matrices.
Here they are:
\begin{mathletters} \begin{eqnarray} &&\alpha_1 \beta_1
G^\mu_{\quad\mu} = 0\, ,\quad \alpha_1 \beta_1 G_{[\kappa\mu]} = 0 \, ;
\\ &&\nonumber\\ &&2i\alpha_1 \beta_2 F_{\alpha\mu}^{\quad\mu} + \alpha_1
\beta_3 \epsilon^{\kappa\tau\mu}_{\quad\alpha} \widetilde
F_{\kappa\tau,\mu} = 0\, ;\\ &&\nonumber\\ &&2i\alpha_1 \beta_3 \widetilde
F_{\alpha\mu}^{\quad\mu} + \alpha_1 \beta_2
\epsilon^{\kappa\tau\mu}_{\quad\alpha} F_{\kappa\tau,\mu} = 0\, ;\\
&&\nonumber\\
&& 2i\alpha_2 \beta_4 T^{\mu}_{\quad\mu\alpha} -
 \alpha_3 \beta_{7}
\epsilon^{\kappa\tau\mu}_{\quad\alpha} \widetilde T_{\kappa,\tau\mu}
= 0\, ;\\
&&\nonumber\\
&& 2i\alpha_3 \beta_{7} \widetilde
T^{\mu}_{\quad\mu\alpha} -
\alpha_2 \beta_4 \epsilon^{\kappa\tau\mu}_{\quad\alpha}
T_{\kappa,\tau\mu} = 0\, ;\\
&&\nonumber\\
&& i\epsilon^{\mu\nu\kappa\tau} \left [ \alpha_2 \beta_6 \widetilde
R_{\kappa\tau,\mu\nu} + \alpha_3 \beta_{8} \widetilde
D_{\kappa\tau,\mu\nu} \right ] + 2\alpha_2 \beta_5
R^{\mu\nu}_{\quad\mu\nu}  + 2\alpha_3
\beta_{9} D^{\mu\nu}_{\quad \mu\nu}  = 0\, ;\\
&&\nonumber\\
&& i\epsilon^{\mu\nu\kappa\tau} \left [ \alpha_2 \beta_5 R_{\kappa\tau,
\mu\nu} + \alpha_3 \beta_{9} D_{\kappa\tau, \mu\nu} \right ]
+ 2\alpha_2 \beta_6 \widetilde R^{\mu\nu}_{\quad\mu\nu}
+ 2\alpha_3 \beta_{8} \widetilde D^{\mu\nu}_{\quad\mu\nu}  =0\, ;\\
&&\nonumber\\
&& 2i \alpha_2 \beta_5 R_{\beta\mu}^{\quad\mu\alpha} + 2i\alpha_3
\beta_{9} D_{\beta\mu}^{\quad\mu\alpha} + \alpha_2 \beta_6
\epsilon^{\nu\alpha}_{\quad\lambda\beta} \widetilde
R^{\lambda\mu}_{\quad\mu\nu} +\alpha_3 \beta_{8}
\epsilon^{\nu\alpha}_{\quad\lambda\beta} \widetilde
D^{\lambda\mu}_{\quad \mu\nu} = 0\, ;\\
&&\nonumber \\
&&2i\alpha_1 \beta_2 F^{\lambda\mu}_{\quad\mu} - 2 i \alpha_2 \beta_4
T_\mu^{\quad\mu\lambda} + \alpha_1 \beta_3 \epsilon^{\kappa\tau\mu\lambda}
\widetilde F_{\kappa\tau,\mu} +\alpha_3 \beta_7
\epsilon^{\kappa\tau\mu\lambda} \widetilde T_{\kappa,\tau\mu} =0\, ;\\
&&\nonumber\\
&&2i\alpha_1 \beta_3 \widetilde F^{\lambda\mu}_{\quad\mu} - 2 i \alpha_3
\beta_7 \widetilde T_\mu^{\quad\mu\lambda} + \alpha_1 \beta_2
\epsilon^{\kappa\tau\mu\lambda} F_{\kappa\tau,\mu} +\alpha_2
\beta_4 \epsilon^{\kappa\tau\mu\lambda}  T_{\kappa,\tau\mu} =0\, ;\\
&&\nonumber\\
&&\alpha_1 \beta_1 (2G^\lambda_{\quad\alpha} - g^\lambda_{\quad\alpha}
G^\mu_{\quad\mu} ) - 2\alpha_2 \beta_5 (2R^{\lambda\mu}_{\quad\mu\alpha}
+2R_{\alpha\mu}^{\quad\mu\lambda} + g^\lambda_{\quad\alpha}
R^{\mu\nu}_{\quad\mu\nu}) +\nonumber\\
&+& 2\alpha_3 \beta_9
(2D^{\lambda\mu}_{\quad\mu\alpha} + 2D_{\alpha\mu}^{\quad\mu\lambda}
+g^\lambda_{\quad\alpha} D^{\mu\nu}_{\quad\mu\nu})+
2i\alpha_3 \beta_8 (\epsilon_{\kappa\alpha}^{\quad\mu\nu}
\widetilde D^{\kappa\lambda}_{\quad\mu\nu} -
\epsilon^{\kappa\tau\mu\lambda} \widetilde D_{\kappa\tau,\mu\alpha}) -
\nonumber\\
&-& 2i\alpha_2 \beta_6 (\epsilon_{\kappa\alpha}^{\quad \mu\nu}
\widetilde R^{\kappa\lambda}_{\quad\mu\nu} -
\epsilon^{\kappa\tau\mu\lambda} \widetilde R_{\kappa\tau,\mu\alpha})
= 0\, ; \\
&&\nonumber\\
&& 2\alpha_3 \beta_8 (2\widetilde D^{\lambda\mu}_{\quad\mu\alpha} + 2
\widetilde D_{\alpha\mu}^{\quad\mu\lambda} +g^\lambda_{\quad\alpha}
\widetilde D^{\mu\nu}_{\quad\mu\nu}) - 2\alpha_2 \beta_6 (2\widetilde
R^{\lambda\mu}_{\quad\mu\alpha} +2 \widetilde
R_{\alpha\mu}^{\quad\mu\lambda} + \nonumber\\
&+& g^\lambda_{\quad\alpha} \widetilde
R^{\mu\nu}_{\quad\mu\nu}) +
2i\alpha_3 \beta_9 (\epsilon_{\kappa\alpha}^{\quad\mu\nu}
D^{\kappa\lambda}_{\quad\mu\nu}  - \epsilon^{\kappa\tau\mu\lambda}
D_{\kappa\tau,\mu\alpha} ) -\nonumber\\
&-& 2i\alpha_2 \beta_5
(\epsilon_{\kappa\alpha}^{\quad\mu\nu} R^{\kappa\lambda}_{\quad\mu\nu}
- \epsilon^{\kappa\tau\mu\lambda} R_{\kappa\tau,\mu\alpha} ) =0\, ;\\
&&\nonumber\\
&&\alpha_1 \beta_2 (F^{\alpha\beta,\lambda} - 2F^{\beta\lambda,\alpha}
+ F^{\beta\mu}_{\quad\mu}\, g^{\lambda\alpha} - F^{\alpha\mu}_{\quad\mu}
\, g^{\lambda\beta} ) - \nonumber\\
&-&\alpha_2 \beta_4 (T^{\lambda,\alpha\beta}
-2T^{\beta,\lambda\alpha} + T_\mu^{\quad\mu\alpha} g^{\lambda\beta} -
T_\mu^{\quad\mu\beta} g^{\lambda\alpha} ) +\nonumber\\
&+&{i\over 2} \alpha_1 \beta_3 (\epsilon^{\kappa\tau\alpha\beta}
\widetilde F_{\kappa\tau}^{\quad\lambda} +
2\epsilon^{\lambda\kappa\alpha\beta} \widetilde F_{\kappa\mu}^{\quad\mu} +
2 \epsilon^{\mu\kappa\alpha\beta} \widetilde F^\lambda_{\quad\kappa,\mu})
-\nonumber\\
&-& {i\over 2} \alpha_3 \beta_7 ( \epsilon^{\mu\nu\alpha\beta} \widetilde
T^{\lambda}_{\quad\mu\nu} +2 \epsilon^{\nu\lambda\alpha\beta} \widetilde
T^\mu_{\quad\mu\nu} +2 \epsilon^{\mu\kappa\alpha\beta} \widetilde
T_{\kappa,\mu}^{\quad\lambda} ) =0\, .
\end{eqnarray}
\end{mathletters}
As one can readily see there is no direct connection between
two $R^{\mu\nu\alpha\beta}$ and two $D^{\mu\nu\alpha\beta}$.
They are only related by formulas containing their various contractions.


\end{document}